\documentclass[doublecol,letterpaper]{epl2}

\usepackage{epsfig}
\usepackage{psfrag}
\usepackage{amsfonts}          
\usepackage{amssymb}           
\usepackage{amsmath}           
\usepackage{ulem}

\newcommand{\dd}{\partial}
\newcommand{\de}{{\rm \, d}}

\renewcommand{\vec}[1]{\mbox{\boldmath $ #1$}}

\usepackage{color}
\renewcommand{\revision}[1]{#1}
\newcommand{\MD}{{\bf MD}}
\newcommand{\FD}{{\bf FD}}
\newcommand{\PA}{$P_\mathrm{MD}$}
\newcommand{\PB}{$P_\mathrm{FD}$}

\newcommand{\mypsfrag}[2]{\psfrag{#1}{\small{#2}}}

\allowdisplaybreaks

\title{Bistability and hysteresis of dipolar dynamos generated by
  turbulent  convection in rotating spherical shells}
\shorttitle{Bistability and hysteresis of dipolar dynamos}

\author{R.~D.~Simitev\inst{1} \and F.~H.~Busse\inst{2}}
\shortauthor{R.~D.~Simitev \& F.~H.~Busse}

\institute{
  \inst{1} Department of Mathematics, University of Glasgow,
  Glasgow G12 8QW, UK\\
  \inst{2} Institute of Physics, University of Bayreuth, Bayreuth
  D-95440, Germany
}
\pacs{91.25.Cw}{Origins and models of the magnetic field; dynamo theories}
\pacs{92.60.hk}{Convection, turbulence, and diffusion}
\pacs{47.20.Ky}{Nonlinearity, bifurcation, and symmetry breaking}

\abstract{
Bistability and hysteresis of magnetohydrodynamic dipolar dynamos
generated by turbulent convection in rotating spherical fluid shells
is demonstrated.  Hysteresis appears as a transition between two
distinct regimes of dipolar dynamos with rather different properties
including a pronounced difference in the amplitude of the axisymmetric
poloidal field component and in the form of the differential
rotation. The bistability occurs from the onset of dynamo action up to
about 9 times the critical value of the Rayleigh number for onset
of convection  and over a wide range of values of the ordinary and the
magnetic Prandtl numbers including the value unity. \\
\textbf{R.~D.~Simitev and F.~H.~Busse, 2009, Europhys. Lett., 85,
  19001\\
  doi:~10.1209/0295-5075/85/19001} }

\begin{document}

\maketitle

\section{Introduction}
An arbitrary weak magnetic field may either decay or be
amplified by the motion of an electrically conducting fluid. The
latter case is called  dynamo effect    
and it is believed to be responsible
for the magnetic fields of cosmic objects including the sun and most planets
\cite{Rüdiger2004Magnetic, DoSo2007}.
The dynamo effect of turbulent convection in rotating spherical
fluid shells has received much attention in recent years  
\cite{Busse2000Homogeneous,Kono2002Recent}
because it is the basic model for the generation of the magnetic fields of the
Earth and other planets. Many
numerical studies following 
\cite{Glatzmaiers1995Threedimensional,Kuang1997Earthlike} 
have attempted a direct comparison with geomagnetic observations and have
been remarkably successful in reproducing some of the main 
properties of the geomagnetic field \cite{DormyNumerical,Ku2004} while others
have focused on more systematic explorations of the computationally accessible parameter
space e.g.~\cite{
Christensen1999Numerical,
Grote2000Regular,
Simitev2005Prandtlnumber,
Taka2005,
Christensen2006Scaling}.  
Dynamos obtained through numerical simulations are usually
turbulent and it is assumed that their time averaged properties are
independent of the initial conditions once the computations have
run for a sufficiently long time.   
In contrast to this assumption we wish to demonstrate in this Letter
that bistability and hysteresis of convection-driven turbulent dynamos
occur in a wide region of the parameter space. The
\revision{co-existence} of two distinct turbulent attractors is also
of general interest. In contrast to the common bistability
\revision{associated with a subcritical bifurcation from a}
laminar to a turbulent attractor, the co-existence of two turbulent
attractors  is a
rare phenomenon in fluid dynamics \cite{Ravelet2004,Mujica2006} and
\revision{magnetohydrodynamics. 
Subcritical onset of dynamo action is a rather common phenomenon
\cite{Busse1977,Ponty2007Subcritical,Kuang2008} and such is hysteresis between non-magnetic and 
dynamo states. But
co-existence and hysteresis between two fully-developed chaotic dynamo
states far above onset are, to our knowledge, reported here for the first time.  
That multiple turbulent states are more likely to occur in hydromagnetic dynamos than in
non-magnetic flows is perhaps not surprising because of the additional degrees of freedom
offered by the magnetic field.} 

\section{Formulation}
We consider a spherical fluid shell of thickness $d$ rotating with a
constant angular velocity $\Omega$. The existence of a static state is
assumed with {a} temperature distribution $T_S = T_0 - \beta d^2 r^2 /2$
and a gravity field in the form $\vec g = - d \gamma \vec r$, where
$rd$ is the length of the position vector with respect to the center
of the sphere. {This form of temperature profile alludes
to the possibility that at least a fraction of the energy available to
planetary dynamos is due to radiogenic heat release.}
In addition to  $d$, we use the time $d^2 / \nu$,  the temperature
$\nu^2 / \gamma \alpha d^4$ and 
the magnetic flux density $\nu ( \mu \varrho )^{1/2} /d$ as scales for
the dimensionless description of the problem  where $\nu$ denotes the
kinematic viscosity of the fluid, $\kappa$ its thermal diffusivity, 
$\varrho$ its density and $\mu$ its magnetic permeability.
In common with most other simulations of Earth and planetary
dynamos \cite{DormyNumerical,Kono2002Recent},
we assume the Boussinesq approximation implying a constant density
$\varrho$ except in the gravity term where its temperature
dependence is taken into account with $\alpha \equiv - (
\de\varrho/\de T)/\varrho =${\sl const}. 
The equations of motion for the velocity vector $\vec u$, the heat
equation for the deviation  $\Theta$ from the static temperature 
distribution, and the equation of induction for the magnetic flux
density $\vec B$ are then given by  
\begin{subequations}
\begin{gather}
\label{1b}
\nabla \cdot \vec u = 0, \qquad \nabla \cdot \vec B = 0, \\
\label{1a} 
(\partial_t + \vec u \cdot \nabla )\vec u + \tau \vec k \times
\vec u = - \nabla \pi +\Theta \vec r + \nabla^2 \vec u + \vec B \cdot
\nabla \vec B, \\
\label{1c}
P(\partial_t \Theta + \vec u \cdot \nabla \Theta) = R \vec r \cdot \vec u + \nabla^2 \Theta, \\
\label{1d} 
\nabla^2 \vec B =  P_m(\partial_t \vec B + \vec u \cdot \nabla \vec B
-  \vec B \cdot \nabla \vec u),
\end{gather}
\end{subequations}
where all gradient terms in the equation of motion 
have been combined into $ \nabla \pi$. The dimensionless
parameters in our formulation are the Rayleigh number $R$, the
Coriolis number $\tau$, the Prandtl number $P$ and the magnetic
Prandtl number $P_m$,  
\begin{equation}
R = \frac{\alpha \gamma \beta d^6}{\nu \kappa} , 
\enspace \tau = \frac{2
\Omega d^2}{\nu} , \enspace P = \frac{\nu}{\kappa} , \enspace P_m = \frac{\nu}{\lambda},
\end{equation}
where $\lambda$ is the magnetic diffusivity.  Being solenoidal vector
fields  $\vec u$ and $\vec B$ can be represented uniquely in terms
of poloidal and toroidal components,
\begin{subequations}
\begin{gather}
\vec u = \nabla \times ( \nabla v \times \vec r) + \nabla w \times 
\vec r \enspace , \\
\vec B = \nabla \times  ( \nabla h \times \vec r) + \nabla g \times 
\vec r \enspace .
\end{gather}
\end{subequations}
We assume fixed temperatures at $r=r_i \equiv 2/3$ and  $r=r_o \equiv
5/3$ and stress-free {rather than no-slip boundary conditions in order
to approach, at least to some extent, the extremely low values of viscosity
believed to be appropriate to planetary cores
\cite{Kuang1997Earthlike}},
\begin{equation}
\label{vbc}
\hspace*{-8mm}
v = \partial^2_{rr}v = \partial_r (w/r) = \Theta = 0.
\end{equation}
For the magnetic field we
assume electrically insulating boundaries at $r=r_i$ and  $r=r_o$ such
that the poloidal function $h$ matches the function $h^{(e)}$ which
describes the potential fields  outside the fluid shell,  
\begin{figure}
\begin{center}
\hspace{-8mm}
\mypsfrag{l}{$l$}
\hspace*{10mm}\epsfig{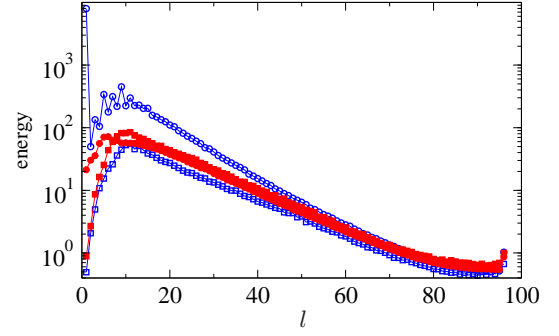}
\end{center}
\caption[]{(color online) Resolution test: Time-averaged spectra of
  magnetic (circles) and kinetic (squares) energy as a function of the
harmonic degree $l$ for the two cases shown in fig.~\ref{f.01}. The \MD{}
case is given in blue and indicated by empty symbols and the \FD{}
case is given in red and indicated by solid symbols.}
\label{f.00}
\end{figure}
\begin{figure}
\begin{center}
\mypsfrag{E}{$E$}
\mypsfrag{M}{$M$}
\mypsfrag{t}{$t$}
\epsfig{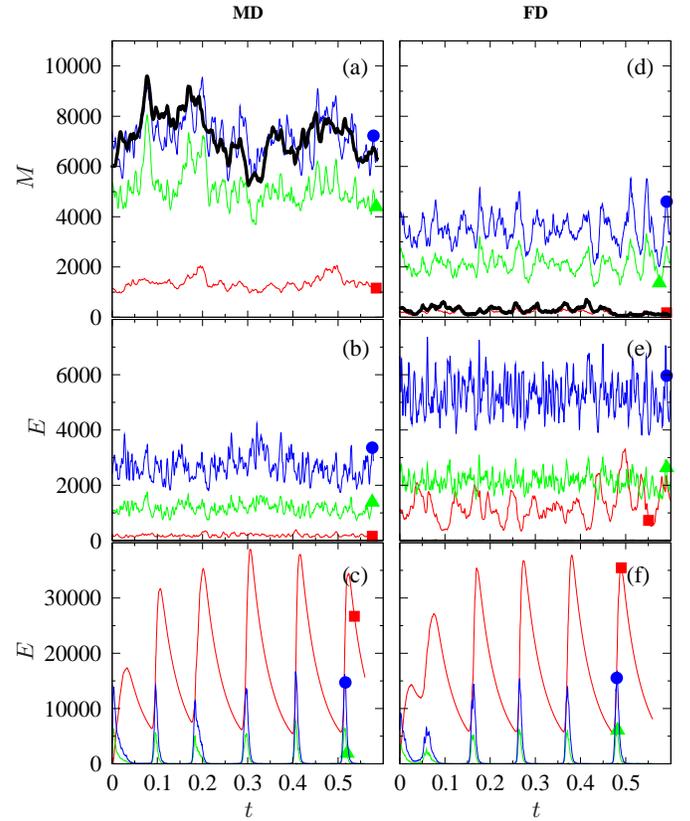}
\end{center}
\caption[]
{(color online) Time series of two different chaotic attractors are
  shown - a \MD{} (left column (a,b)) and a 
\FD{} dynamo (right column (d,e)) both in the case $R=3.5\times10^6$,
$\tau=3\times10^4$,
  $P=0.75$ and  $P_m=1.5$.
The top two panels (a,d) show magnetic energy densities.
The rest of the panels show kinetic energy densities in the presence
of magnetic field (b,e) and after the magnetic field is removed (c,f).
The component $\overline{X}_p$ is shown by thick solid black line, while
$\overline{X}_t$, $\widetilde{X}_p$, and $\widetilde{X}_t$ are shown
by thin red, green and blue lines, respectively, and
they are also indicated by squares, triangles and circles,
respectively. $X$ stands for either $M$ or $E$.
}
\label{f.01}
\end{figure}
\begin{figure}
\begin{center}
\mypsfrag{MD}{{\scriptsize\MD}}
\mypsfrag{FD}{{\scriptsize\FD}}
\epsfig{file=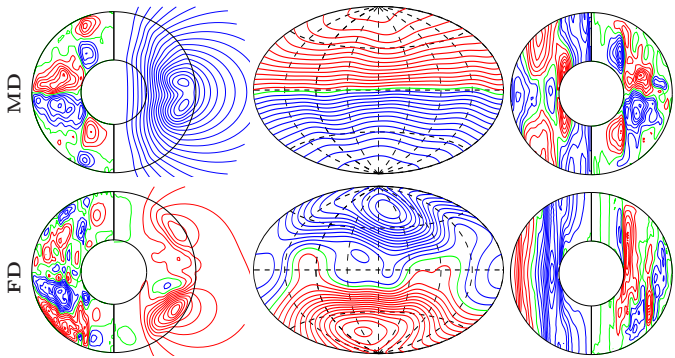,width=\columnwidth,clip=}
\end{center}
\caption[]{(color online) A \MD{} and a \FD{} dynamo both in the case $R=3.5\times10^6$,
  $\tau=3\times10^4$, $P=0.75$ and $P_m=1.5$. Each circle of the leftmost
  column shows meridional lines of constant $\overline{B_{\varphi}}$
  in the left half and of $r \sin \theta \dd_\theta
  \overline{h}=const.$ in the right half. The middle column
  shows lines of constant $B_r$ at $r=r_o+1.3$. Each circle of the rightmost
  column shows meridional lines of constant $\overline{u}_\varphi$ in
  the left half and of $r \sin \theta \dd_\theta \overline{v}$ in the
  right half.
}
\label{f.04}
\end{figure}
%
%
\begin{equation}
\hspace*{-8mm}
\label{mbc}
g = h-h^{(e)} = \partial_r ( h-h^{(e)})=0\;\; \mbox{at}\;\; r=r_i, r_o.
\end{equation}
The radius ratio $r_i/r_o = 0.4$ is slightly larger than that
appropriate for the Earth's liquid core. 
{This is a standard formulation of the spherical
  convection-driven dynamo problem
  \cite{DoSo2007,Busse2000Homogeneous,Kono2002Recent} for which an extensive
  collection of results already exists
  \cite{Grote2000Regular,GroBu,Simitev2005Prandtlnumber,Busse2006Parameter}. 
The results reported below are not strongly model dependent. In
particular, dynamos with stress-free and with no-slip velocity boundary
conditions as well as with different modes of energy supply are known to
have comparable energy densities and symmetry properties (see fig.~15 of
\cite{Kono2002Recent}). 
Furthermore, aiming to retain a general physical perspective, we
intentionally use a minimal number of physical parameters including only
those of primary  importance for stellar and planetary applications.}

Equations of motion for the scalar fields $v$, $w$, are obtained by
taking $\vec r\cdot\nabla\times\nabla\times$~ and $\vec
r\cdot\nabla\times$~   of equation \eqref{1a} and
equations for $g$ and $h$ are obtained by taking $\vec
r\cdot\nabla\times$~ and $\vec r\cdot$~  of equation \eqref{1d}. 
These equations are solved numerically by a 
pseudo-spectral method as described in \cite{Tilgner1999Spectral} 
based on expansions of all dependent variables in
spherical harmonics for the angular dependences and in Chebychev
polynomials for the radial dependence. 
{Typically, calculations are considered decently  resolved
  when the spectral power of kinetic and magnetic energy drops by more
  than a factor of 100 from the spectral maximum to the cut-off
  wavelength \cite{Christensen1999Numerical}.} A minimum of 41
collocation 
points in the radial direction and spherical harmonics up to the
order 96 have been used in all cases reported here {which provides
adequate resolution as demonstrated in fig.~\ref{f.00} for two
typical dynamo solutions.} 
The dynamo solutions are characterized by their magnetic energy
densities, 
\begin{gather}
\overline{M}_p = \frac{1}{2} \langle \mid\nabla \times ( \nabla \overline{h}
\times \vec r )\mid^2 \rangle ,  \quad
 \overline{M}_t = \frac{1}{2} \langle \mid\nabla
\overline g \times \vec r \mid^2 \rangle, \nonumber\\
\widetilde{M}_p = \frac{1}{2} \langle \mid\nabla \times ( \nabla \widetilde h
\times \vec r )\mid^2 \rangle , \quad 
 \widetilde{M}_t = \frac{1}{2} \langle \mid\nabla
\widetilde g \times
\vec r \mid^2 \rangle, \nonumber
\end{gather}
where $\langle\cdot\rangle$ indicates the average over the fluid shell
and $\overline h$ refers to the axisymmetric component of $h$,
while $\widetilde h$ is defined by $\widetilde h = h - \overline h $. 
The corresponding kinetic energy densities $\overline{E}_p$,
$\overline{E}_t$, $\widetilde{E}_p$ and $\widetilde{E}_t$
are defined analogously with $v$ and $w$ replacing $h$ and
$g$. 
%
%
%
%
%
\begin{figure}[t]
\begin{center}
\mypsfrag{0.3}    {0.3}    
\mypsfrag{0.5}    {0.5}    
\mypsfrag{0.75}   {0.75}   
\mypsfrag{1}      {1}      
\mypsfrag{2}      {2}      
\mypsfrag{3}      {3}      
\mypsfrag{5}      {5}      
\mypsfrag{10}     {10}     
\mypsfrag{0.15}   {0.15}   
\mypsfrag{0.2}    {0.2}    
\mypsfrag{0.3}    {0.3}    
\mypsfrag{0.4}    {0.4}    
\mypsfrag{0.5}    {}    
\mypsfrag{0.7}    {0.7}    
\mypsfrag{1}      {1}      
\mypsfrag{10}     {10}     
\mypsfrag{0}      {0}      
\mypsfrag{10}     {10}     
\mypsfrag{1}      {1}      
\mypsfrag{P}      {$P$}      
\mypsfrag{P/Pm}   {$P/P_m$}   
\mypsfrag{Mpf/Mpm}{$\widetilde{M}_p/\overline{M}_p$}
\hspace*{-2mm}
\epsfig{file=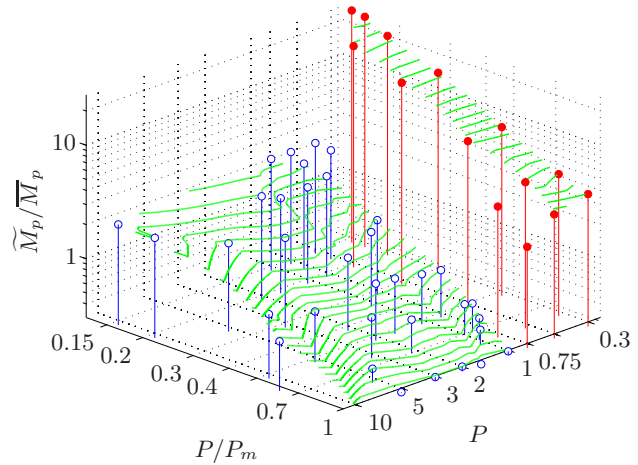,width=\columnwidth,clip=}
\end{center}
\caption[]{(color online) The ratio $\widetilde{M}_p/\overline{M}_p$ of dynamo
  solutions as a function of $P$ and $P/P_m$ at $R=3.5\times10^6$,
  $\tau=3\times10^4$. 
  Full red and empty blue circles indicate \FD{} and \MD{} dynamos, respectively. The
  surface is intentionally left broken near the transition where it is
  multi-valued as discussed in text.} 
\label{f.02}
\end{figure}
\begin{figure*}
\begin{center}
\mypsfrag{0.1}{0.1}
\mypsfrag{1}{1}
\mypsfrag{10}{10}
\mypsfrag{0.1}{0.1}
\mypsfrag{0.3}{0.3}
\mypsfrag{0.5}{0.5}
\mypsfrag{3}{3}
\mypsfrag{0.2}{0.2}
\mypsfrag{0.4}{0.4}
\mypsfrag{0}{0}
\mypsfrag{1}{1}
\mypsfrag{2}{2}
\mypsfrag{3}{3}
\mypsfrag{4}{4}
\mypsfrag{5}{5}
\mypsfrag{6}{6}
\mypsfrag{7a}{$7\times10^6$}
\mypsfrag{P}{$P$}
\mypsfrag{R}{$R$}
\mypsfrag{PPm}{\hspace*{-2mm}$P/P_m$}
\mypsfrag{ratio}{$\widetilde{M}_p/\overline{M}_p$}
\mypsfrag{Nui}{$\mathrm{Nu_i}$}
\mypsfrag{a}{(a)}
\mypsfrag{b}{(b)}
\mypsfrag{c}{(c)}
\mypsfrag{PA}{\PA}
\mypsfrag{PB}{\PB}
\mypsfrag{sA}{}
\hspace*{-7mm}
\epsfig{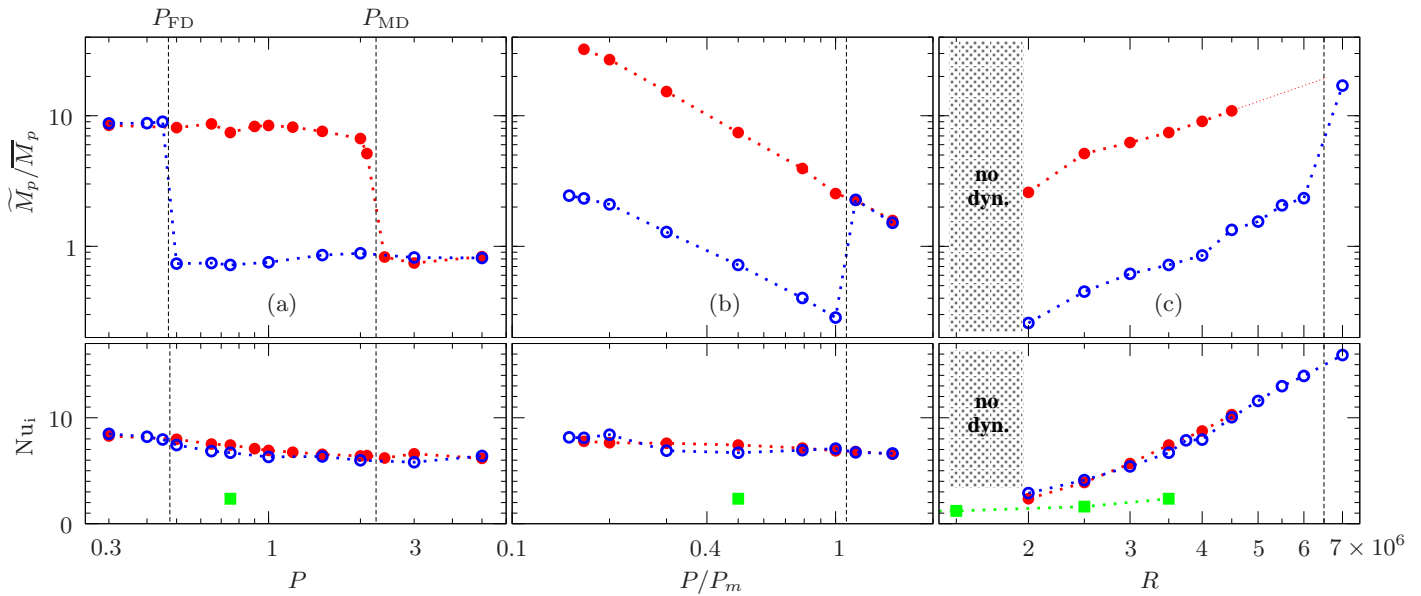}
\end{center}
\caption[]{(color online) The upper row shows the hysteresis effect in the ratio
  $\widetilde{M}_p/\overline{M}_p$ at $\tau=3\times10^4$ 
  (a) as a function of the Prandtl number in the case 
  of $R=3.5\times10^6$, $P/P_m=0.5$; (b) as a function of
  the ratio  $P/P_m$ in the case of $R=3.5\times10^6$,
  $P=0.75$ and (c)  as a function of the Rayleigh number in the case
  $P=0.75$, $P_m=1.5$. Full red and empty
  blue circles indicate \FD{} and   \MD{} dynamos, respectively. 
The critical value of $R$ for the onset of
  thermal convection for the cases shown in (c) is $R_c=659145$. \revision{A
  transition from \FD{} to \MD{} dynamos as $P/P_m$ is decreased in
  (b) is expected, but is not indicated owing to lack of data.}
  The lower row shows the value of the Nusselt number at
  $r=r_i$ for the same dynamo cases. Values for non-magnetic convection are
  indicated by green  squares for comparison. }
\label{f.03}
\end{figure*}

\section{Results}
All solutions reported below are turbulent and typical examples of
the variations in time of the energy densities are shown in fig.~\ref{f.01}.  
Apart from the obvious quantitative difference, an
essential qualitative change in the balance of magnetic energy
components is observed. The axisymmetric poloidal component 
$\overline{M}_p$  is dominant in the case shown in
fig.~\ref{f.01}(a,b) while it has a relatively small contribution in 
the case of fig.~\ref{f.01}(d,e) with the corresponding differences
in the structure of the magnetic field shown in fig.~\ref{f.04}. 
This observation is in accord with the claim made in
\cite{Kutzner2002From,Simitev2005Prandtlnumber,Busse2006Parameter} that, quite
generally, two regimes of dipolar dynamos can be distinguished, namely
those with $\widetilde{M}_p < \overline{M}_p$ (regime \MD, "Mean
Dipole") and those with $\widetilde{M}_p > \overline{M}_p$ (regime \FD,
"Fluctuating Dipole").
The transition between the regimes is shown in fig.~\ref{f.02} as a
function of the ordinary Prandtl number $P$ and of the ratio $P/P_m$
in the case of fixed $\tau=3\times10^4$ and $R=3.5\times10^6$.  
The transition between the two distinct turbulent dynamo attractors is
not gradual, but has the character of an 
abrupt jump after a critical parameter value is surpassed as discussed
below. 
All solutions included in fig.~\ref{f.02} have a predominantly
dipolar character with the ratio $\overline{M}_p^{\mathrm{dip}}/\overline{M}_p$ in
the ranges $[0.95, 1]$ for \MD{} dynamos and $[0.45, 0.72]$ for \FD{}
dynamos. 
Although, the \FD{} dynamos feature an increased contribution of
higher multipoles they are still of  geophysical relevance
\cite{Kutzner2002From,Busse2006Parameter}. At values of the ordinary 
and magnetic Prandtl numbers somewhat lower than those included in
fig.~\ref{f.02}, however, hemispherical and quadrupolar dynamos become predominant.   

We note that the two solutions in fig.~\ref{f.01} and fig.~\ref{f.04}
are obtained at identical parameter values and thus the chaotic attractors \MD{} and \FD{}
co-exist in this case. Varying $P$ and $P_m$
demonstrates an extended region of this co-existence.  In fact, 
the transition between the \MD{} and \FD{} dynamos is achieved
via hysteresis loops as illustrated in fig.~\ref{f.03} for
$\tau=3\times10^4$.                   
When an \MD{} dynamo is used
as initial data and the Prandtl number $P$ is decreased, solutions
remain in regime \MD{} until the critical value \PB$\approx0.5$ is reached at
which point an abrupt jump transition to the \FD{} regime occurs. 
When a \FD{} dynamo is used as initial condition and $P$ is
increased the reverse transition occurs at the critical value
\PA$\approx2.2$ as seen in
fig.~\ref{f.03}(a). Similarly, coexisting attractors are 
found as a function of the ratio $P/P_m$ for $P/P_m < 1$
and as a function of the Rayleigh number above the onset of dynamo
action as seen in figs.~\ref{f.03}(b) and (c), respectively. The
solutions of the hysteresis loops have been computed for up to at
least 3 magnetic diffusion times. No evidence for a transient nature
of any case has been found. In fact, in cases beyond the boundaries of
the hysteresis loop the transition
from \MD{} to \FD{} regimes or vice versa takes typically only 0.15
magnetic diffusion times. 
\revision{One may notice that the hysteresis loop in the direction of
decreasing values of $P/P_m$ in fig.~\ref{f.03}(b) is incomplete. We
expect a transition from \FD{} to \MD{} dynamos to occur at sufficiently low values of 
$P/P_m$, but we are presently unable to demonstrate it because of the high
computational costs of dynamo simulations in this region of the parameter space.}

The bistability of convection driven dynamos is the result of two
different ways in which the magnetic field damps the differential
rotation. In the \MD{} dynamos the differential rotation generated by
Reynolds stresses of the convection columns is eliminated almost
entirely by the strong mean magnetic field as shown in
figs.~\ref{f.01}(b) and \ref{f.04}. Only a zonal thermal wind caused by
latitudinal variations of the temperature 
remains as is typical for high Prandtl number dynamos
\cite{Simitev2005Prandtlnumber}. Because of the strong magnetic field
the amplitude of convection is also reduced in comparison with the
maximum value that it reaches in the absence of a magnetic field. In
the case of \FD{} dynamos the differential rotation is still 
diminished, but its alignment with coaxial cylindrical surfaces is
preserved. The amplitude of convection is now more strongly
fluctuating, but is larger on average than in the case of the \MD{}
dynamos. In this way \FD{} and \MD{} dynamos manage to carry very nearly
the same heat transport as is evident from fig.~\ref{f.03}. This heat
transport by far exceeds the time average of heat transport found in
the absence of a magnetic field. Figure \ref{f.01}(c,f) demonstrates that the
same type of convection in the form of turbulent relaxation
oscillations \cite{GroBu} is generated when the Lorentz force is
dropped from the equations of motion.   
\begin{figure}
\begin{center}
\epsfig{file=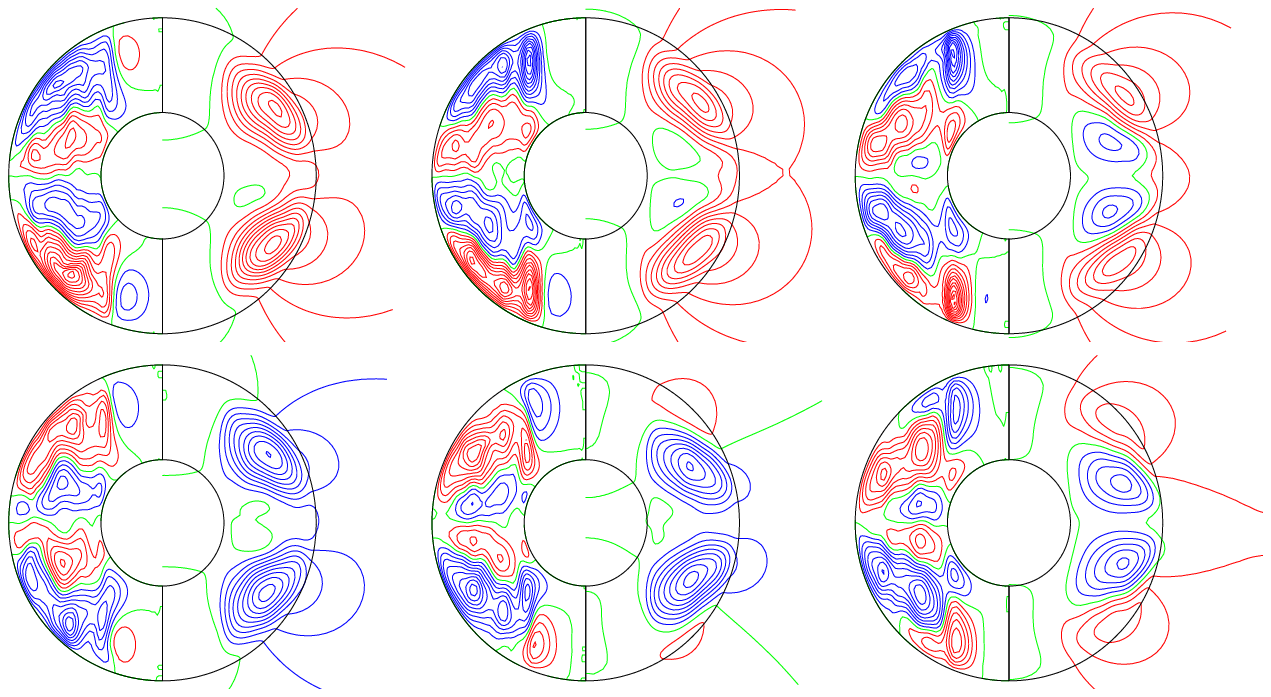,width=\columnwidth,clip=}
\end{center}
\caption[]{(color online) A half period of dipolar oscillations in the \FD{} case 
  $R=3.5\times10^6$, $\tau=3\times10^4$, $P=0.75$ and $P_m=0.65$. The same
  fields as in the first column of fig.~\ref{f.04} are shown. 
  Plots  follow clockwise from upper left with a step $\Delta t=0.0042$.}
\label{f.05}
\end{figure}

While the \MD{} dynamos are non-oscillatory with limited fluctuations
of their axisymmetric components about their time average, the \FD{}
dynamos usually oscillate in an irregular manner. 
A rather pure example of such oscillations is shown in
fig.~\ref{f.05}. The dipolar oscillations of \FD{} dynamos can be
understood  in terms of Parker dynamo waves
\cite{Busse2006Parameter}. In general, however, the oscillations are
much less regular and involve considerable contributions of
quadrupolar components. 

\section{Conclusion}
We have demonstrated the co-existence of two well-distinguishable chaotic
attractors in a turbulent system over a sizable region of the parameter
space. The hysteresis loops occur within the ranges $P\in(0.5,2.2)$ and
$P/P_m<1$ for the value of $\tau$ used here and from onset of
dynamo action to at least $9\times R_c$, where $R_c$ is the critical
value for onset of convection. This range is of importance since $P=1$
is used in most simulations of convection-driven dynamos in rotating
spheres \cite{DormyNumerical,Kono2002Recent}. The reason for this is
that in comparisons with geomagnetic observations diffusivities such
as $\nu$ and  $\kappa$ are interpreted as eddy diffusivities which are
supposed to represent the effects of the numerically unresolved scales
of the turbulent velocity field. Since the diffusion of heat and
momentum owing to the small scale turbulence is similar, eddy
diffusivities are regarded as equal with the consequence $P=1$.  

\revision{Dynamo states similar to \MD{} and \FD{} dynamos have been
obtained previously. A transition between such two states as a function of
the Rayleigh number, $R$, has been} reported in \cite{Kutzner2002From}
where no-slip boundary conditions have been used. A hysteresis phenomenon
was not found by these authors. In a later paper
\cite{Christensen2006Scaling} a particular case of bistability has
been mentioned, but no further discussion was given. 
\revision{Transitions between states similar to \MD{} and
\FD{} dynamos as a function of $P$ or $P/P_m$ caused by the changing
strength of the inertial forces 
have been reported in \cite{Simitev2005Prandtlnumber,Sreenivasan2006}.
In the simulations of \cite{Kutzner2002From,Christensen2006Scaling} and
\cite{Sreenivasan2006} no-slip boundary conditions for the
velocity field have been employed and it is of  interest to investigate
whether, as we expect, the hysteresis phenomenon persist as the velocity boundary
conditions and the heating model are changed. Simulations of
convection-driven dynamos with no-slip boundary conditions and driven
by a basal heat flux are underway and will be reported in a future 
paper.}

While the possibility of hysteretic behavior  of planetary dynamos in
response to slow changes in their  properties is of considerable
interest, it will be difficult to obtain observational evidence for
such a  phenomenon because of the long time scale of magnetic
diffusion. It is important, however, to be aware of the bistability
phenomenon in the interpretation of numerical {dynamo} simulations.

\bibliographystyle{eplbib}
\bibliography{rs}
\end{document}